\documentclass[conference]{IEEEtran}
\IEEEoverridecommandlockouts
\usepackage{cite}
\usepackage{amsmath,amssymb,amsfonts}
\usepackage{algorithmic}
\usepackage{graphicx}
\usepackage{textcomp}
\usepackage{float}
\usepackage{xcolor}
\usepackage{scalerel}
\usepackage{tikz}
\usepackage{background}
\def\BibTeX{{\rm B\kern-.05em{\sc i\kern-.025em b}\kern-.08em
    T\kern-.1667em\lower.7ex\hbox{E}\kern-.125emX}}

\backgroundsetup{
  position=current page.south,
  angle=1,
  scale=1,
  contents={}
}

\AddEverypageHook{%
  \ifnum\value{page}=1
    \backgroundsetup{
      scale=1,
      color=black,
      opacity=1,
      angle=90,
      position={current page.west},
      vshift=-0.8cm,
      hshift=-0.1cm,
      contents={\footnotesize
        2024 IEEE International Symposium on Circuits and Systems (ISCAS) $|$ 979-8-3503-3099-1/24/$\$$31.00 ©2024 IEEE $|$ DOI: 10.1109/ISCAS58744.2024.10558539%
      }
    }
    \BgMaterial%
  \fi
}

\newcommand{\addfootnote}{
    \begin{tikzpicture}[remember picture,overlay]
        \node[anchor=south,yshift=0.5cm] at (current page.south) {
            \parbox{\textwidth}{\centering\footnotesize
            \textcopyright 2024 IEEE.  Personal use of this material is permitted.  Permission from IEEE must be obtained for all other uses, in any current or future media, including reprinting/republishing this material for advertising or promotional purposes, creating new collective works, for resale or redistribution to servers or lists, or reuse of any copyrighted component of this work in other works.
            }
        };
    \end{tikzpicture}
}

\begin{document}

\title{Assessing the Performance of Stateful Logic in 1-Selector-1-RRAM Crossbar Arrays
}

\author{\IEEEauthorblockN{Arjun Tyagi}
\IEEEauthorblockA{\textit{Electrical and Computer Engineering} \\
\textit{Technion - Israel Institue of Technology}\\
Haifa, Israel \\
arjun.tyagi@campus.technion.ac.il}
\and
\IEEEauthorblockN{Shahar Kvatinsky}
\IEEEauthorblockA{\textit{Electrical and Computer Engineering} \\
\textit{Technion - Israel Institue of Technology}\\
Haifa, Israel \\
shahar@ee.technion.ac.il}
}

\maketitle

\addfootnote

\begin{abstract}
Resistive Random Access Memory (RRAM) crossbar arrays are an attractive memory structure for emerging nonvolatile memory due to their high density and excellent scalability. Their ability to perform logic operations using RRAM devices makes them a critical component in non-von Neumann processing-in-memory architectures. Passive RRAM crossbar arrays (1-RRAM or 1R), however, suffer from a major issue of sneak path currents, leading to a lower readout margin and increasing write failures. To address this challenge, active RRAM arrays have been proposed, which incorporate a selector device in each memory cell (termed 1-selector-1-RRAM or 1S1R). The selector eliminates currents from unselected cells and therefore effectively mitigates the sneak path phenomenon. Yet, there is a need for a comprehensive analysis of 1S1R arrays, particularly concerning in-memory computation. In this paper, we introduce a 1S1R model tailored to a VO\textsubscript{2}-based selector and TiN/TiO\textsubscript{x}/HfO\textsubscript{x}/Pt RRAM device. We also present simulations of 1S1R arrays, incorporating all parasitic parameters, across a range of array sizes from $4\times4$ to $512\times512$. We evaluate the performance of Memristor-Aided Logic (MAGIC) gates in terms of switching delay, power consumption, and readout margin, and provide a comparative evaluation with passive 1R arrays.

\end{abstract}


\section{Introduction}
\label{introduction}
The resistive random access memory (RRAM) based on metal-oxide has been regarded as a promising option for the next-generation nonvolatile memory application due to its advantageous characteristics such as low programming voltage ($<$ 3 V), fast switching speed ($<$ 10 ns), excellent scalability ($<$ 10 nm), and compatibility with CMOS fabrication~\cite{akinaga2010resistive, wong2012metal}. RRAM typically consists of a simple metal-insulator-metal structure with two terminals. Its operation relies on the voltage-driven resistance switching of the metal-oxide between a low-resistance state (LRS) and a high-resistance state (HRS). The process of transitioning from HRS to LRS is known as the SET process, while the reverse is referred to as the RESET process. In bipolar RRAM, the SET process is achieved by applying a positive SET voltage (V\textsubscript{set}), and the RESET process is achieved by applying a negative RESET voltage (V\textsubscript{reset}).

\begin{figure}[t]
    \vspace*{-10pt}
    \centering
    \includegraphics[width=0.80\linewidth]{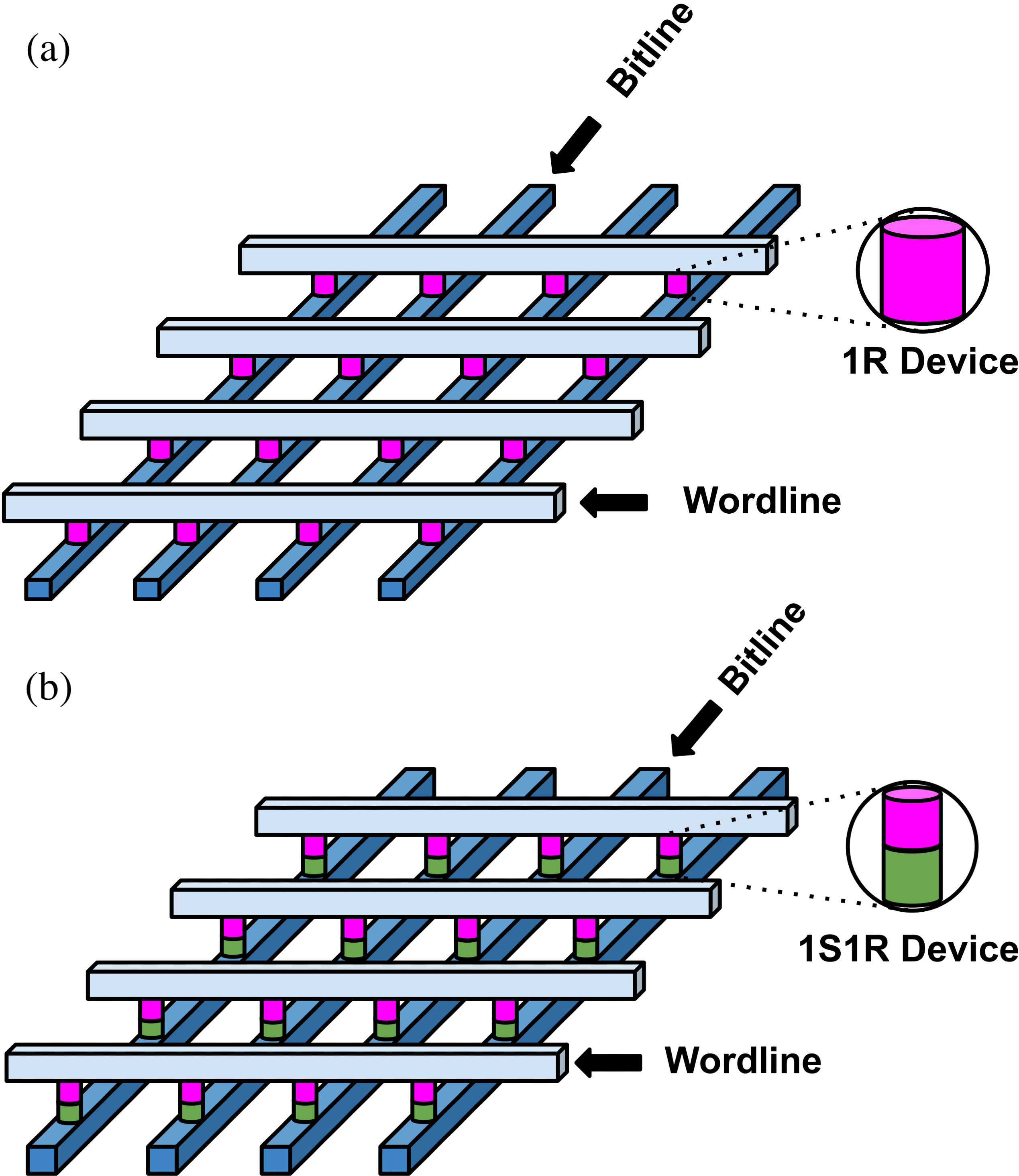}
    \caption{(a) Passive RRAM or 1R crossbar array, and (b) active RRAM or 1S1R crossbar array. The selector and RRAM device are illustrated in green and pink respectively.}
    \label{crossbar_array_architecture}
    \vspace{-20pt}
\end{figure}

Generally, an RRAM array can be built in a 1-RRAM (1R) crossbar array, also known as passive RRAM crossbar array, as illustrated in Fig.~\ref{crossbar_array_architecture}(a). An RRAM cell is located at the intersection of each bitline (BL) and wordline (WL). The top electrode of the RRAM is connected to the BL while the bottom electrode is connected to the WL.
This architecture, however, suffers from the sneak path current problem~\cite{burr2014access, jiang2016compact, linn2010complementary, cassuto2016information}, where current flows through unselected memory cells, changing the current sensed during read operations. This phenomenon also increases the probability for write disturbance. Another memory architecture to alleviate the sneak path current is the 1-selector-1-RRAM (1S1R) or active RRAM crossbar array which has a two-terminal selector in series with an RRAM device, together making up a 1S1R cell as shown in Fig. \ref{crossbar_array_architecture}(b). The top terminal of the selector is connected to the BL while the bottom electrode of RRAM device is connected to the WL. The selector can effectively cut off the leakage current at low-voltage bias, and thus eliminate the sneak path currents.

Various types of selector devices such as Ovonic Threshold Switching (OTS) and Insulator-Metal Transition (IMT) have been explored and integrated with RRAM devices to create 1S1R cells. Robayo \textit{et al.} proposed the integration of Ge-Sb-Se-N (GSSN) OTS selector with HfO\textsubscript{2}/Ti RRAM, thereby suppressing the leakage current to less than 1 nA \cite{robayo2019integration}. Son \textit{et al.} presented a nanoscale VO\textsubscript{2} selector device with excellent characteristics like fast switching speed ($<$ 20 ns) and high current density ($>$ 10\textsuperscript{6} A/cm\textsuperscript{2}), resulting in a significantly improved readout margin when integrated with ZrO\textsubscript{x}/HfO\textsubscript{x} bipolar RRAM device~\cite{son2011excellent}. However, while most of these works focused on memory applications, very few have studied the proposed 1S1R cell in crossbar arrangement for in-memory computing operations. 
This paper focuses on studying the 1S1R crossbar array for in-memory computation. We design a compact Verilog-A model for 1S1R cell and fit it to a VO\textsubscript{2}-based selector \cite{son2011excellent} and TiN/TiO\textsubscript{x}/HfO\textsubscript{x}/Pt RRAM device~\cite{jiang2016compact, farjadian2022modeling}. To accurately simulate crossbar array, we model parasitic resistances and capacitance that play a critical role in determining the performance of a crossbar array~\cite{wald2019understanding}. We perform extensive simulation on both 1S1R and 1R crossbar arrayss of varying sizes ($4\times4$ to $512\times512$) for Memristor-Aided Logic (MAGIC), a popular stateful logic family, and analyze the performance in terms of switching delay, power consumption, and readout margin. Our results indicate that, on average, 1S1R crossbar array consumes $21.9\times$ and $24.4\times$ less power than 1R crossbar array for NOR(0,0) and NOR(0,1)/NOR(1,0)/NOR(1,1) operations, respectively. Furthermore, the readout margin of 1S1R crossbar array is, on average, $4.5\times$ more than 1R crossbar array, thereby enabling much larger crossbar arrays.

This paper is arranged as follows: Section \ref{background} provides a brief background on Memristor-Aided loGIC (MAGIC). The Verilog-A model for 1S1R and parameter fitting are discussed in Section \ref{model}. We discuss the setup of crossbar array simulation in Section \ref{crossbar_array_simulation} and present the results in Section \ref{results}. Finally, Section \ref{conclusion} concludes the paper.

\begin{figure}
    \vspace*{-10pt}
    \centering
    \includegraphics[width=0.99\linewidth]{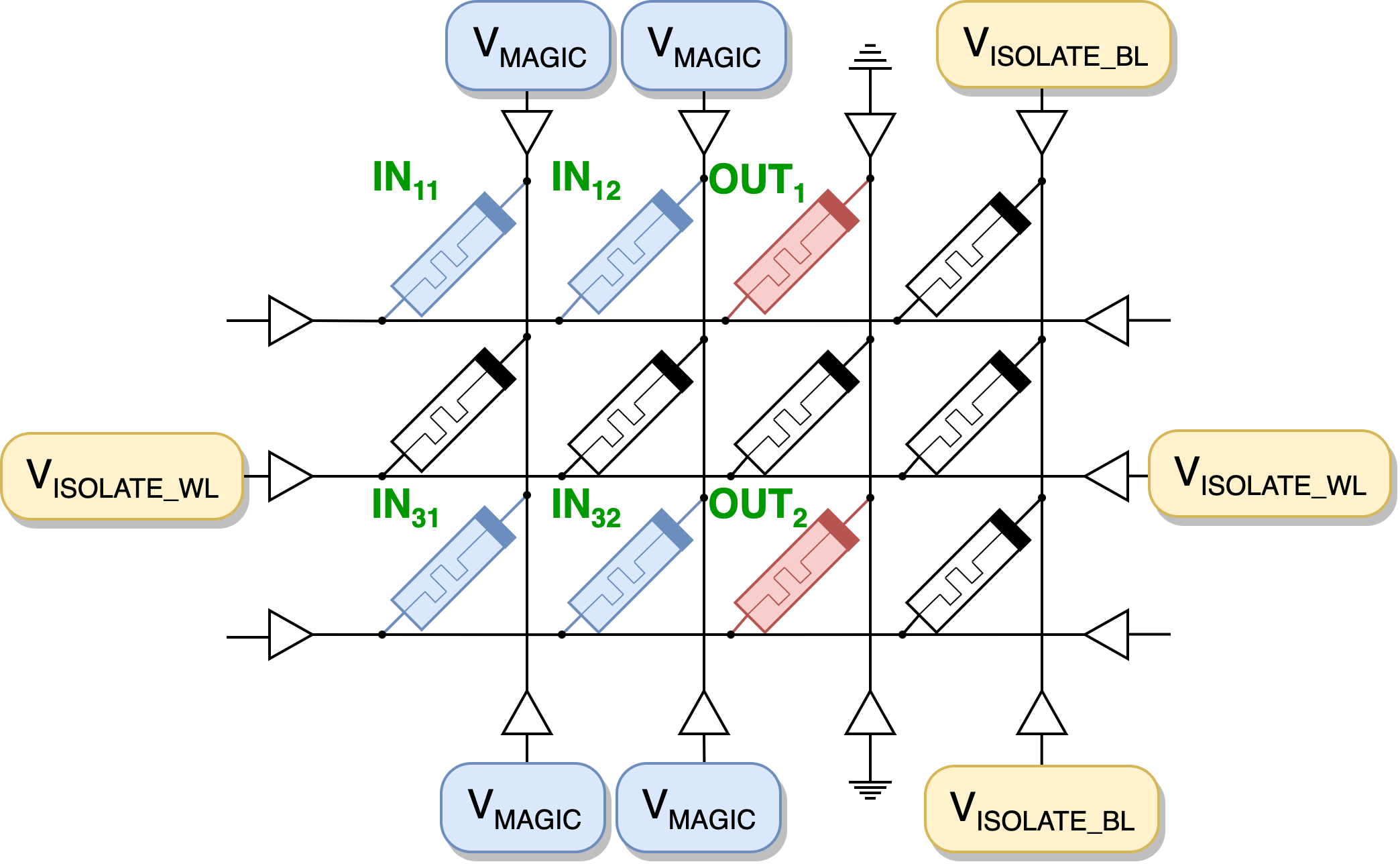}
    \caption{MAGIC NOR gates performed over column vectors.}
    \label{MAGIC}
     \vspace{-10pt}
\end{figure}

\section{Memristor Aided Logic (MAGIC)}
\label{background}
MAGIC~\cite{MAGIC} is a stateful logic family suitable for processing-in-memory using RRAM. A MAGIC gate requires three voltages (V\textsubscript{MAGIC}, GND, and V\textsubscript{ISOLATE}) to perform a NOR logic operation. The inputs and output of the MAGIC gate are the resistance of memristors, similarly to the representation of data in the memristive memory. As NOR is a functionally complete operation, MAGIC NOR is sufficient to execute any Boolean function~\cite{Logic_Design_MAGIC}. Due to the structure and symmetry of the crossbar array, MAGIC operations can be performed across multiple rows or columns simultaneously. Fig. \ref{MAGIC} shows the schematic of a MAGIC NOR operation performed over multiple rows. While V\textsubscript{ISOLATE} mitigates the sneak path during computation, it does not eliminate it completely. Furthermore, passive crossbar array still exhibit sneak path currents during memory operations, as required in MAGIC architectures.

\section{1S1R Model}
\label{model}
This section describes the selector and RRAM models that together make up the 1S1R model.

\subsection{Selector Model}
Insulator-Metal Transition (IMT)-based selector switches its state depending on the voltage across it. Being a symmetric bipolar selector, it switches from OFF state to ON state when the voltage across it exceeds (subceeds) a threshold voltage $V_{th}$ ($-V_{th}$). The selector switches back to OFF state when the voltage across it subceeds (exceeds) a voltage $V_{hold}$ ($-V_{hold}$), wherein 0 $< V_{hold} < V_{th}$.

In the OFF state, the selector follows the Poole-Frenkel conduction model~\cite{farjadian2022modeling}, whereas in ON state, the selector can be approximated using Ohm's Law. Therefore, the I-V relationship of the selector can be modelled as:

\begin{equation}
  I = \begin{cases}
    \frac{V}{\beta_{\scaleto{s}{3pt}}} \times \exp({\frac{V - V_{\scaleto{s}{3pt}}}{\alpha_{\scaleto{s}{3pt}}}}), & \text{OFF state}
    \\[3mm]
    \frac{V}{R_{\scaleto{ON}{3pt}}}, & \text{ON state}\\
    
  \end{cases},
\end{equation}
where $I$ and $V$ are the current through the selector and the voltage across it, respectively, $R_{ON}$ is the resistance of the selector in ON state, and $\alpha_{s}$, $\beta_{s}$ and $V_{s}$ are fitting parameters. 

\begin{table}[b]
    \vspace{-20pt}
    \caption{Fitting parameters for 1S1R model.}
    \vspace{-5pt}
    \label{1S1R_Fitting_Parameters}
    \centering
    \includegraphics[width=0.9\linewidth]{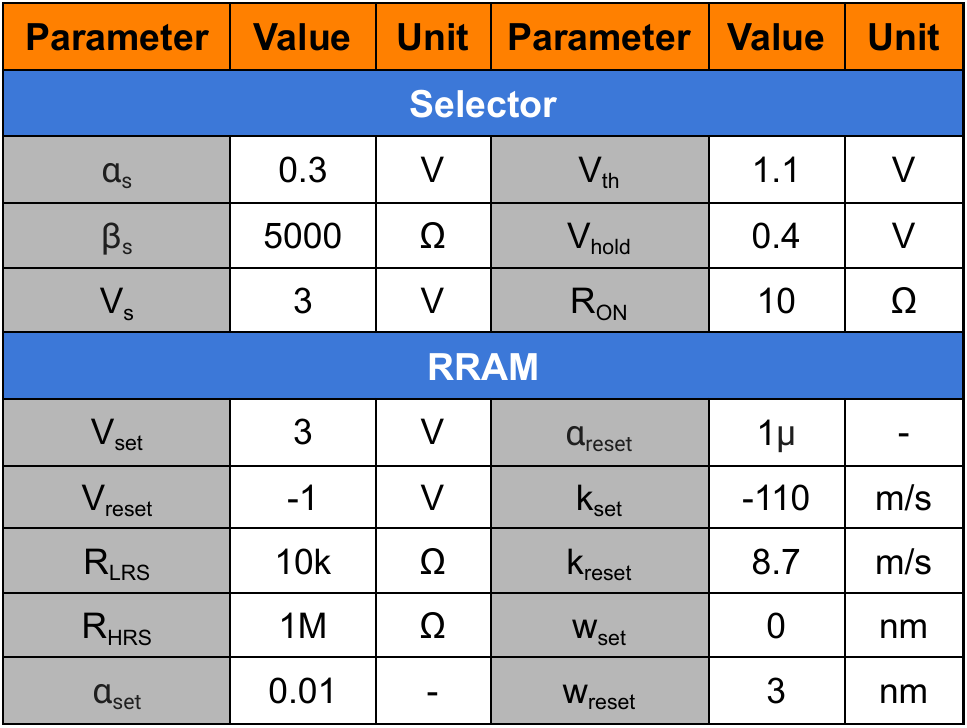}
    \vspace{-10pt}
\end{table}

\subsection{RRAM Model}
For the RRAM model, we use the VTEAM model \cite{kvatinsky2015vteam} albeit with minor changes\footnote{In our model implementation, $V_{set}$ and $V_{reset}$ are, respectively, positive and negative voltages.}. The VTEAM model is based on an expression of the derivative of the internal state variable as follows,

\vspace{-5pt}
\begin{equation}
    \frac{dw(t)}{dt} = \begin{cases}
        k_{\scaleto{reset}{3pt}} \cdot (\frac{V}{V_{\scaleto{reset}{3pt}}} - 1) ^{\alpha_{\scaleto{reset}{3pt}}} \cdot f_{\scaleto{reset}{3pt}}(w), & \text{if V $<$ V\textsubscript{reset}},\\
        
        k_{\scaleto{set}{3pt}} \cdot (\frac{V}{V_{\scaleto{set}{3pt}}} - 1) ^{\alpha_{\scaleto{set}{3pt}}} \cdot f_{\scaleto{set}{3pt}}(w), & \text{if V $>$ V\textsubscript{set}},\\

        0, & \text{otherwise}, \\
    \end{cases}
\end{equation}

where $k_{reset}$, $k_{set}$, $\alpha_{reset}$, and $\alpha_{set}$ are constants and $V_{reset}$ and $V_{set}$ are the RESET and SET voltage, respectively. Functions $f_{\scaleto{reset}{3pt}}(w)$ and $f_{\scaleto{set}{3pt}}(w)$ represent the dependence of the derivative of the state variable on state variable $w$.

The I-V relationship for the RRAM follows an exponential behavior as:

\begin{equation}
    I = \frac{e^{-\frac{\lambda}{w_{\scaleto{reset}{2.5pt}} - w_{\scaleto{set}{2.5pt}}} \cdot (w - w_{\scaleto{set}{2.5pt}})}}{R_{\scaleto{LRS}{3pt}}} \cdot V,
\end{equation}
where $w_{reset}$ and $w_{set}$ are the bounds of the internal state variable $w$. $R_{LRS}$ is the resistance of the device at low resistance state and $\lambda$ is the fitting parameter such that $e^\lambda = \frac{R_{\scaleto{HRS}{3pt}}}{R_{\scaleto{LRS}{3pt}}}$.

\begin{figure}[t] 
    \vspace{-10pt}
    \centering
    \includegraphics[width=0.95\linewidth]{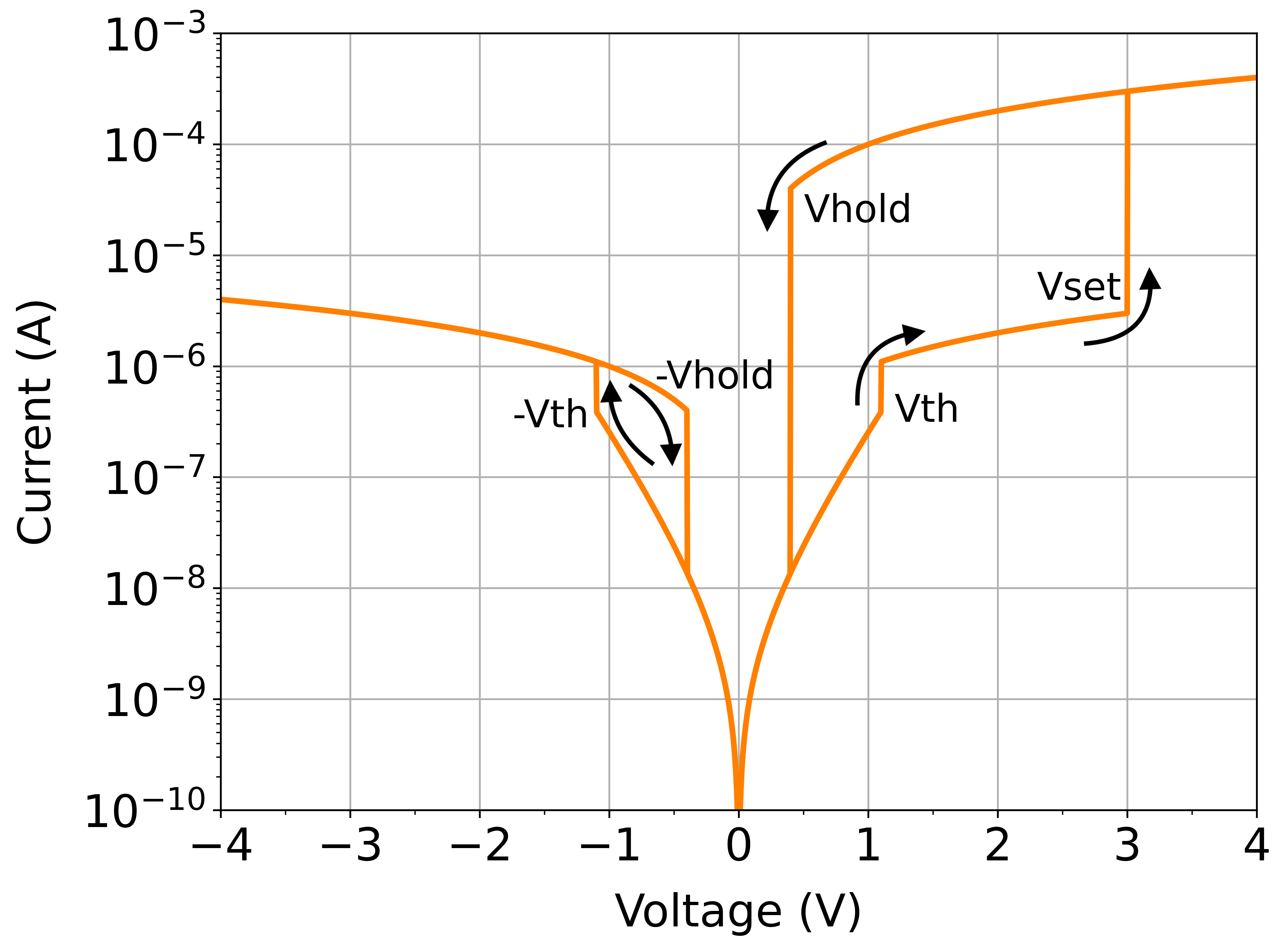}
    \vspace{-10pt}
    \caption{I-V characteristics of the 1S1R cell. The cell is in HRS for forward sweep and LRS for reverse sweep.}
    \label{1S1R_IV_Characteristics}
    \vspace{-10pt}
\end{figure}

\subsection{Parameter Fitting}
We fit the 1S1R model to a VO\textsubscript{2}-based selector \cite{son2011excellent} and TiN/TiO\textsubscript{x}/HfO\textsubscript{x}/Pt RRAM device \cite{jiang2016compact} by varying the fitting parameters of the model such that we match the model to the actual switching delay of $1.2ns$ and $6.6ns$ \cite{farjadian2022modeling} for SET and RESET operation, respectively. We choose $V_{set}$ and $V_{reset}$ as $3V$ and $-1V$, respectively, to meet the switching criterion to correctly perform MAGIC operation. Table \ref{1S1R_Fitting_Parameters} lists the fitting parameters of the model and fig. \ref{1S1R_IV_Characteristics} shows the I-V characteristic of the 1S1R cell during forward and reverse sweep.

\section{Crossbar Array Simulations}
\label{crossbar_array_simulation}
\begin{figure}[t]
    \vspace{-10pt}
    \centering
    \includegraphics[width=0.95\linewidth]{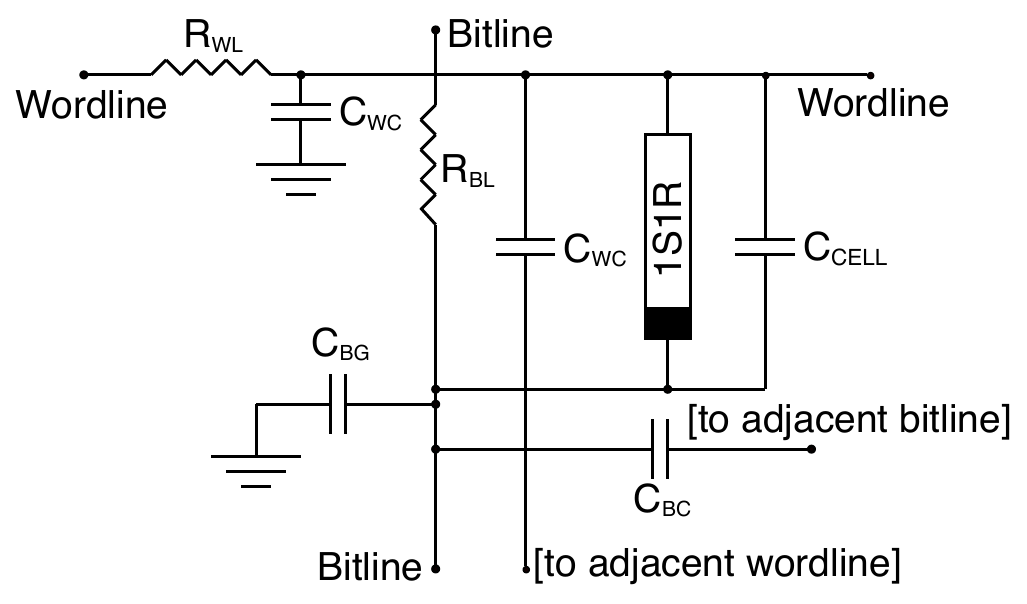}
    \vspace{-10pt}
    \caption{A 1S1R cell comprising of parasitic resistances and capacitance~\cite{9531867}.}
    \label{1S1R_cell}
    \vspace{-10pt}
\end{figure}

To study the performance of the 1S1R cell, we design several symmetric crossbar arrays of different sizes in the Cadence Design Suite. Since a crossbar array consists of several interconnects running parallel to each other, it leads to parasitic resistances and capacitances that can significantly degrade the performance of the crossbar array. Therefore, it is important to incorporate these parasitics for crossbar array simulation. Fig. \ref{1S1R_cell} shows the schematic of the unit cell used for crossbar array simulation including the parasitics~\cite{9531867}.

The output cell of the MAGIC gate, for both 1R and 1S1R, is at the center of the crossbar array to account for worst case scenario. The two input cells are taken to be adjacent to the output cell. To compare the performance of 1S1R crossbar arrays, we also simulate 1R crossbar arrays of same sizes with the same RRAM device model.


\begin{figure}[b]
    \vspace{-15pt}
    \centering
    \includegraphics[width=0.98\linewidth]{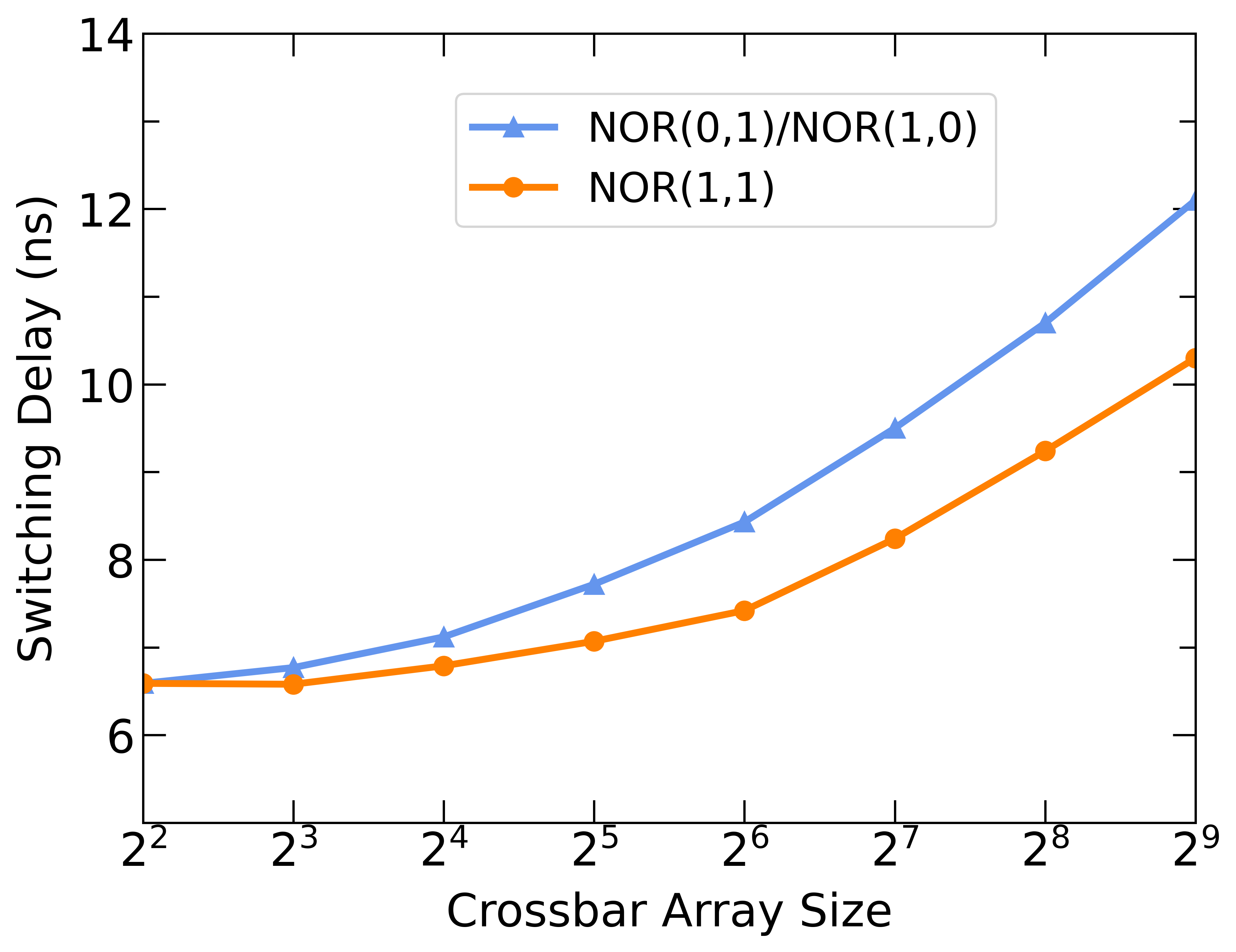}
    \vspace{-10pt}
    \caption{Switching delay for NOR(0,1)/NOR(1,0) and NOR(1,1) operation for different array sizes.}
    \label{Switching_Delay}
    \vspace{-15pt}
\end{figure}

\section{Results}
\label{results}

\begin{figure*}[t]
    \vspace{-10pt}
    \centering
    \includegraphics[width=0.79\textwidth]{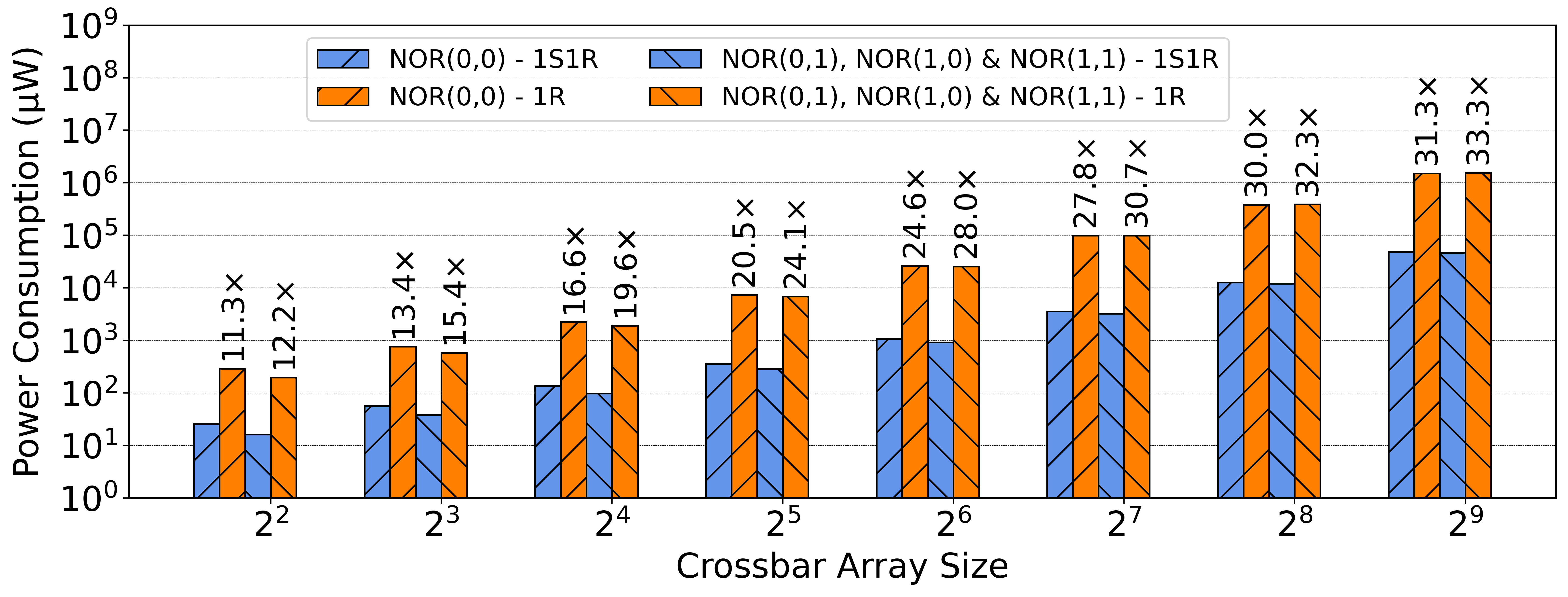}
    \vspace*{-8pt}
    \caption{Power consumption for NOR(0,0) and NOR(0,1)/NOR(1,0)/NOR(1,1) operations on different 1R and 1S1R crossbar array sizes; (lower is better). }
    \label{Power_Consumption}
    \vspace{-5pt}
\end{figure*}

\begin{figure}[b]
    \vspace{-10pt}
    \centering
    \includegraphics[width=0.89\linewidth]{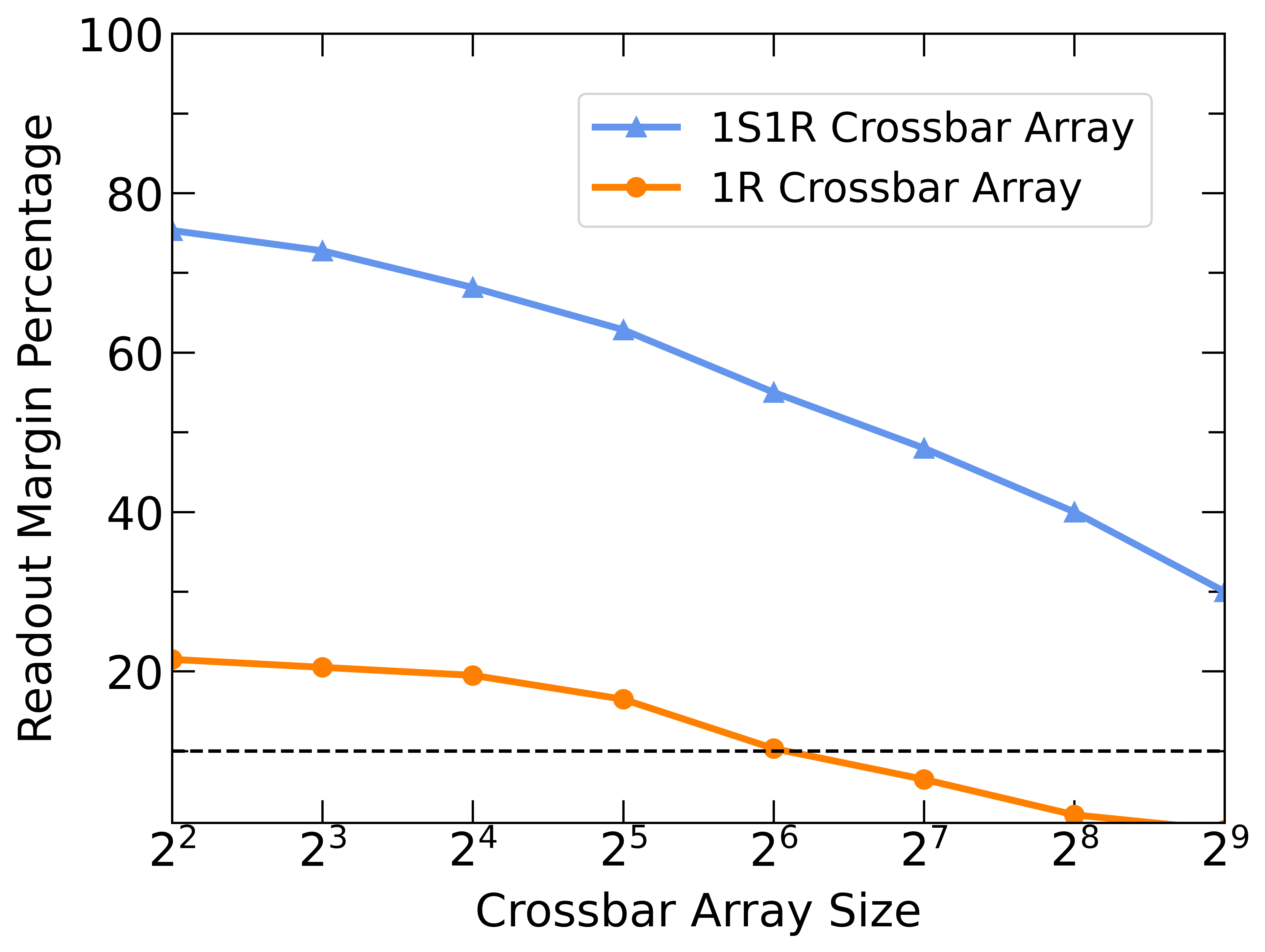}
    \vspace*{-8pt}
    \caption{Readout margin as percentage of read voltage for different array sizes.}
    \label{Readout_Margin}
    \vspace{-15pt}
\end{figure}

To evaluate the performance of 1S1R crossbar array, we simulate array sizes of $4\times4$ to $512\times512$.
We focus on three parameters: (1) switching delay at the output 1S1R cell for NOR(0,1)/NOR(1,0) and NOR(1,1) operations, (2) power consumed to perform a MAGIC NOR operation with all input combinations, and (3) readout margin of the crossbar array when performing READ operation. To measure switching delay and power consumption, we apply $V_{MAGIC}$ of $3V$ for $20ns$. We choose $V_{ISOLATE\_WL}$ and $V_{ISOLATE\_BL}$ as $1V$ and $2V$, respectively, to isolate the rows and columns not involved in MAGIC operations. For power consumption measurements, we activate half of the total rows in each crossbar array. For readout margin, we apply a $V_{READ}$ voltage of $2V$ at the selected bitline and connected the selected wordline to $R_{SENSE} = \sqrt{R_{LRS}\times R_{HRS}} = 100k\Omega$ \cite{zhou2014crossbar}. For the unselected bitlines and wordlines, we employ 1/3 V scheme wherein unselected bitlines are biased at 2/3 $V_{READ}$ and unselected wordlines are biased at 1/3 $V_{READ}$ \cite{zhou2014crossbar}.

\subsection{Switching Delay}
Fig. \ref{Switching_Delay} shows the switching delay at the output 1S1R cell for NOR(0,1)/NOR(1,0) and NOR(1,1) operations (NOR(0,0) does not lead to switching of output cell). We draw two conclusions from this figure.

First, irrespective of the input combination, the switching delay increases with an increase in crossbar array size. This is because larger crossbar array has more parasitic resistance and capacitance, thereby requiring more time to charge the RC networks. Furthermore, due to the increasing voltage drop across the RC networks, the voltage across the input cells and output cell also decreases for larger crossbar array, leading to higher switching delay.

Second, the switching delay for NOR(1,1) is lower as compared to NOR(0,1)/NOR(1,0). This is because the current through the output cell for NOR(1,1) is slightly higher than for NOR(0,1)/NOR(1,0). The higher current is because of the lower resistance of the input combination for NOR(1,1) ($5k\Omega$) compared to that of NOR(0,1)/NOR(1,0) ($\approx 10k\Omega$).

\subsection{Power Consumption}
For the sake of brevity, we classify MAGIC NOR operation into two parts: (1) neither input is in LRS, and (2) at least one input is in LRS. Fig. \ref{Power_Consumption} shows the power consumed in performing NOR operation for both parts.

Irrespective of the input combination, the ratio of the power consumed by 1R to 1S1R crossbar array increases for larger crossbar arrays. This can be attributed to the fact as with increasing array size, the number of isolated cells increases in both crossbar arrays. While the current through these isolated cells in 1R crossbar array depends on their state (we assume half of them are in HRS and half of them in LRS), for 1S1R crossbar array, the selector device of all the isolated cells is in OFF state, resulting in significantly lower current  and consequently, less power consumption as compared to 1R crossbar array.

\subsection{Readout Margin}
The readout margin \cite{zhou2014crossbar} for a crossbar array is defined as:
\begin{equation}
    RM = \frac{\Delta V_{\scaleto{OUT}{4pt}}}{V_{\scaleto{READ}{4pt}}} = \frac{V_{\scaleto{OUT}{4pt}}(LRS) - V_{\scaleto{OUT}{4pt}}(HRS)}{V_{\scaleto{READ}{4pt}}},
\end{equation}
where $V_{\scaleto{OUT}{4pt}}(LRS)$ and $V_{\scaleto{OUT}{4pt}}(HRS)$ are the output voltages measured across $R_{SENSE}$ when sensing logical 1 (LRS) and logical 0 (HRS), respectively, and $V_{READ}$ is the read voltage applied at the selected bitline. When sensing LRS (HRS), we assume all the unselected cells to be in HRS (LRS). This helps us best capture the effect of sneak path current on the readout margin. Fig. \ref{Readout_Margin} shows the readout margin as a percentage of the read voltage for different crossbar array sizes. We draw two conclusions from the figure.

First, the readout margin is lower for larger crossbar arrays for both 1S1R and 1R crossbar arrays. This is because with increasing array size, there are more unselected cells leading to an increased number of sneak paths. This leads to an increase in $V_{\scaleto{OUT}{4pt}}(HRS)$ for both 1R and 1S1R crossbar array and thus, a decrease in readout margin. 

Second, the readout margin of 1R crossbar array is significantly lower than that of 1S1R crossbar array and goes below the $10\%$ threshold (required for correctly distinguishing between LRS and HRS)~\cite{zhou2014crossbar, upadhyay2021engineering} for array size greater than $2^6$ wordlines. This is because for 1R crossbar array, a significant portion of the total current flows through the unselected cells due to opposite logical state. This issue is aggravated when sensing HRS since all the unselected cells are in LRS, thereby allowing the current to prefer the path of least resistance. On the other hand, for 1S1R crossbar array, irrespective of the logical state of the unselected cells, the selector device for all the unselected cells remains in the OFF state, thereby forcing the current to flow through the selected cell only. This leads to a much higher $V_{\scaleto{OUT}{4pt}}(HRS)$ for 1R than for 1S1R crossbar array, leading to substantial decrease in the readout margin for the 1R crossbar array.

\section{Conclusion}
\label{conclusion}
This paper presents comprehensive simulation of 1S1R crossbar array for MAGIC operations. We present a 1S1R model and fit it to a VO\textsubscript{2}-based selector and TiN/TiO\textsubscript{x}/HfO\textsubscript{x}/Pt RRAM device. We included the parasitic resistance and capacitance of a crossbar array in the unit cell and simulate crossbar arrays from size $4\times4$ to $512\times512$. We assess the performance of 1S1R crossbar array in terms of switching delay, power consumption, and readout margin. Our results indicate that on average, a 1S1R crossbar array consumes $21.9\times$ and $24.4\times$ lower power than 1R crossbar arrays for NOR(0,0) and NOR(0,1)/NOR(1,0)/NOR(1,1) operations, respectively. Furthermore, the readout margin of 1S1R crossbar array is, on average, $4.5\times$ more than 1R crossbar array, thereby enabling much larger crossbar arrays.

\section{Acknowledgment}
\label{Acknowledgment}
This work was supported by the European Research Council through the European Union’s Horizon 2020 Research and Innovation Programme under Grant 757259 and by NSF-BSF grant number 2020-613.

\bibliographystyle{ieeetr}
\bibliography{main.bib}

\end{document}